\documentclass{emulateapj}

\shorttitle{Propagation effects in hot magnetized plasma}
\begin{document}

\title{Propagation effects in magnetized transrelativistic plasmas}
\author{Roman V. Shcherbakov}
\email{rshcherbakov@cfa.harvard.edu} \affil{Harvard-Smithsonian Center for Astrophysics, 60 Garden
Street, Cambridge, MA 02138}

\begin{abstract}
The transfer of polarized radiation in magnetized and non-magnetized relativistic plasmas is an
area of research with numerous flaws and gaps. The present paper is aimed at filling some gaps and
eliminating the flaws. Starting from a Trubnikov's linear response tensor for a vacuum wave with
${\bf k}=\omega/c$ in thermal plasma, the analytic expression for the dielectric tensor is found in
the limit of high frequencies. The Faraday rotation and Faraday conversion measures are computed in
their first orders in the ratio of the cyclotron frequency $\Omega_0$ to the observed frequency
$\omega$. The computed temperature dependencies of propagation effects bridge the known
non-relativistic and ultra-relativistic limiting formulas. The fitting expressions are found for
high temperatures, where the higher orders in $\Omega_0/\omega$ cannot be neglected. The plasma
eigenmodes are found to become linearly polarized at much larger temperatures than thought before.
The results are applied to the diagnostics of the hot ISM, hot accretion flows, and jets.
\end{abstract}

\section{Introduction}
We learn much of our information about astrophysical objects by observing the light they emit.
Observations of the polarization properties of light can tell us the geometry of the emitter,
strength of the magnetic field, density of plasma, and temperature. The proper and correct theory
of optical activity is essential for making accurate predictions. While the low-temperature
propagation characteristics of plasma are well-established \citep{landau10}, the theory of
relativistic effects has not been fully studied. In this paper I discuss the propagation effects
through a homogeneous magnetized relativistic plasma. A non-magnetized case emerges as a limit of
the magnetized case. The discussion is divided into three separate topics.

Two linear plasma propagation effects are Faraday rotation and Faraday conversion \citep{mueller}.
Traditionally, these effects are considered in their lowest orders in the ratio $\beta$ of the
cyclotron frequency $\Omega_0$ to the circular frequency of light $\omega,$ id est in a
high-frequency approximation. The distribution of particles is taken to be thermal
\begin{equation}\label{distrib}
dN=\frac{n \exp (-\gamma/T)}{4\pi m^2 T^2 K_2(T^{-1})}d^3p
\end{equation}
with the dimensionless temperature $T$ in the units of particle rest mass temperature $m c^2/k_B.$
The Faraday rotation measure $RM$ and conversion measure are known in a non-relativistic $T\ll 1$
and an ultra-relativistic $T\gg 1$ limits \citep{melrosec}. I derive a surprisingly simple analytic
expression for arbitrary temperature $T.$

The smallness of $\beta=\Omega_0/\omega,$  $\beta\ll1$ in the real systems led some authors
\citep{melrosea} to conclude that the high-frequency approximation will always work. However, there
is a clear indication that it breaks down at high temperatures $T\gg 1.$ It was claimed that the
eigenmodes of plasma are linearly polarized for high temperatures $T\gg 1$ \citep{melrosec},
because the second order term $\sim\beta^2$ becomes larger than the first order term $\sim \beta$
due to the $T$ dependence. The arbitrarily large $T$-factor may stand in front of higher order
expansion terms in $\beta$ of the relevant expressions. I find the generalized rotation measure as
a function of $\beta$ and $T$ without expanding in $\beta$ and compare the results with the known
high-frequency expressions. The high-$T$ behavior of the plasma response is indeed significantly
different.

Plasma physics involves complicated calculations. This led to a number of errors in the literature
\citep{melrosec}, some of which have still not been fixed. In the article I check all the limiting
cases numerically and analytically and expound all the steps of derivations. Thus I correct the
relevant errors and misinterpretations made by previous authors, hopefully not making new mistakes.
The analytical and numerical results are obtained in Mathematica 6 system. It has an enormous
potential in these problems \citep{mathem_blog}.

The paper is organized as follows. The formalism of plasma response and calculations are described
in \S \ref{method}. Several applications to observations can be found in \S \ref{applications}. I
conclude in \S \ref{discussion} with a short summary and future prospects.

\section{Calculations}\label{method}
\subsection{Geometry of the problem}
\begin{figure}[h]
\plotone{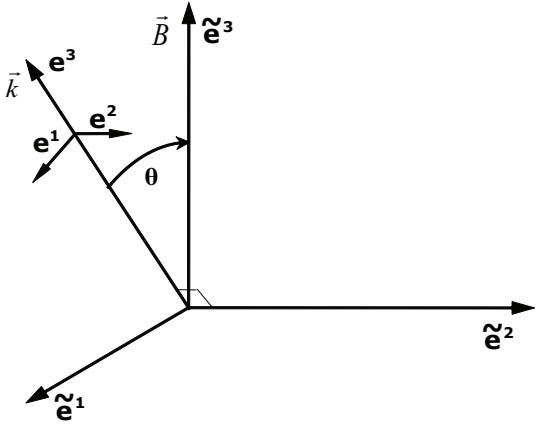} \epsscale{1}
 \caption{Geometry of the problem.}
 \label{fig_rot}
\end{figure}

I assume the traditional geometry depicted on Figure~\ref{fig_rot}: \begin{itemize}
\item{Euclidean basis (${\bf \tilde{e}}^1,$ ${\bf \tilde{e}}^2,$ ${\bf \tilde{e}}^3$),}
\item{magnetic field along the third axis ${\bf \tilde{B}}=(0, 0, B)^T,$} \item{a wave vector of
the wave ${\bf \tilde{k}}=k(\sin\theta,0,\cos\theta)^T$ with an angle $\theta$ between ${\bf
\tilde{k}}$ and ${\bf \tilde{B}}.$}
\end{itemize}
The basis is rotated from  (${\bf \tilde{e}}^1,$ ${\bf \tilde{e}}^2,$ ${\bf \tilde{e}}^3$) to
(${\bf e}^1,$ ${\bf e}^2,$ ${\bf e}^3$), so that the wave propagates along ${\bf k}=(0, 0, k)^T$ in
the new basis. The transformation has the form
\begin{equation}
{\bf e}^1={\bf \tilde{e}}^1 \cos\theta - {\bf \tilde{e}}^3  \sin\theta,\quad {\bf e}^2={\bf
\tilde{e}}^2, \quad {\bf e}^3={\bf \tilde{e}}^1 \sin\theta + {\bf \tilde{e}}^3 \cos\theta,
\end{equation}
which can be conveniently written as
\begin{equation}
{\bf e}^\mu={\bf \tilde{e}}^\nu S^{\nu\mu}, \quad S^{\nu\mu}=\left(\begin{array}{ccc}
\cos\theta & 0 & \sin\theta \\
0 & 1 & 0\\
-\sin\theta & 0 & \cos\theta
\end{array}     \right).
\end{equation}
Vectors  and tensors then rotate according to
\begin{equation}\label{transform}
A^\mu=(S^T)^{\mu\nu}\tilde{A}^\nu, \quad
\alpha^{\mu\nu}=(S^T)^{\mu\sigma}\tilde{\alpha}^{\sigma\delta}S^{\delta \nu}.
\end{equation}

\subsection{Linear plasma response}
The propagation of weak electromagnetic (EM) waves in a homogeneous magnetized plasma can be fully
described by the response tensor $\alpha^{\mu\nu}.$ It expresses the linear proportionality between
the induced current density and the vector potential $j^\mu(\omega)={\alpha^\mu}_\nu
A^{\nu}(\omega).$ The spatial projection of such defined 4-D tensor ${\alpha^\mu}_\nu$ is equal to
the 3-D tensor $\alpha_{i j}$ defined by ${\bf j}=\alpha_{ij}{\bf A}.$

I consider Trubnikov's form of the response tensor \citep{trubnikov, melrosea}. I work in a
low-density regime, where the plasma response is calculated for a vacuum wave with $|{\bf
k}|=\omega/c.$ I take the tensor $\tilde{\alpha}^{\mu\nu}$ from the first-hand derivations
\citep{trubnikov, melrosea}, make the transformation (\ref{transform}), and take the 1-st and 2-nd
components in both indices. Thus the projection onto the (${\bf e}^1, {\bf e}^2$) plane in CGS
units is
\begin{equation}\label{main}
{\alpha^\mu}_\nu(k)=\frac{i q^2 n \omega \rho^2}{c m K_2(\rho)}\int^\infty_0
d\xi\left[{t^\mu}_\nu\frac{K_2(r)}{r^2}- R^\mu\bar{R}_\nu\frac{K_3(r)}{r^3}\right],
\end{equation}
\begin{equation}\label{dt}
{t^\mu}_\nu=\ \left(\begin{array}{cc}
\cos^2\theta\cos\Omega_0\xi+\sin^2\theta & \eta\cos\theta\sin\Omega_0\xi \\
-\eta\cos\theta\sin\Omega_0\xi & \cos\Omega_0\xi
\end{array}     \right) ,
\end{equation}

\begin{mathletters}
\begin{equation}\label{rmu}
R^\mu =\frac{\omega\sin\theta}{\Omega_0}\left(\cos\theta(\sin\Omega_0\xi-\Omega_0\xi),
-\eta(1-\cos\Omega_0\xi)\right),
\end{equation}
\begin{equation}\label{rnu}
\bar{R}_\nu=\frac{\omega\sin\theta}{\Omega_0}\left(\cos\theta(\sin\Omega_0\xi-\Omega_0\xi),
\eta(1-\cos\Omega_0\xi)\right),
\end{equation}
\end{mathletters}
 and

\begin{equation}
r=\left[\rho^2-2i
\omega\xi\rho+\frac{\omega^2\sin^2\theta}{\Omega_0^2}\left(2-\Omega_0^2\xi^2-2\cos\Omega_0\xi\right)\right]^{1/2},
\end{equation}
where $\eta$ is the sign of the charge, $K_n(r)$ is the n-th Bessel function of the second
kind\footnote{Note that the analogous expression in \cite{melrosec} has an extra factor
$\Omega_0\xi$ in the component $t^{11}$ and the opposite sign of $R^\mu\bar{R}^\nu$ term by an
error. The author has corrected his formulas in \cite{melrose_qmag}.}. The quantity $\rho$ is the
dimensionless inverse temperature,
\begin{equation}\rho=T^{-1}=\frac{m c^2}{k_B T_p},\end{equation} where $T_p$ the actual temperature of particles.
 The response of plasma is usually characterized by the dielectric tensor. Its projection onto the (${\bf e}^1,
{\bf e}^2$) plane is
\begin{equation}\label{dielectric}
{\varepsilon^\mu}_{\nu}={\delta^\mu}_\nu+\frac{4\pi c}{\omega^2}{\alpha^\mu}_\nu.
\end{equation}  The wave equation for transverse waves in terms of
${\varepsilon^\mu}_{\nu}$ is
\begin{equation}\label{dispersion}
(n_{\rm r}^2 {\delta^\mu}_\nu - {\varepsilon^\mu}_\nu) \left(\begin{array}{c} E_1 \\ E_2
\end{array}\right)=0,
\end{equation} where $E_1$ and $E_2$ are the components of the electric field along ${\bf e}^1$ and ${\bf e}^2$ and
$n_{\rm r}^2=k^2 c^2/\omega^2$ \citep{swanson}.

\subsection{High frequency limit}\label{high_freq}
Let me first calculate the limiting expression for ${\alpha^\mu}_\nu$ in the high-frequency limit
$\Omega_0\ll\omega.$ I denote
\begin{equation}\label{podst}
\alpha=\omega\xi, \qquad \beta=\frac{\Omega_0}{ \omega},
\end{equation} substitute the definitions (\ref{podst}) into the expression (\ref{main}), and expand the response tensor ${\alpha^\mu}_\nu$ in $\beta$.
I retain only up to the 2-nd order of the expansion, which gives the conventional generalized
Faraday rotation \citep{melrosec}. The first terms of the series of $r,$ ${t^\mu}_\nu,$ and $R^\mu
\bar{R}_\nu$ read
\begin{equation}\label{rad}
r^2= r_0^2+\delta r^2, \quad r_0^2=\rho^2-2 i \alpha \rho, \quad \delta
r^2=-\frac{\sin^2\theta}{12}\beta^2\alpha^4,
\end{equation}
\begin{equation}\label{exp_t}
{t^\mu}_\nu=\left(\begin{array}{cc}
1-\cos^2\theta\cdot\alpha^2\beta^2/2& \alpha\beta\eta\cos\theta \\
-\alpha\beta\eta\cos\theta & 1-\alpha^2\beta^2/2
\end{array}     \right) ,
\end{equation}

\begin{equation}\label{exp_R}
R^\mu \bar{R}_\nu=-\frac{\alpha^4 \beta^2}4\sin^2\theta\left(\begin{array}{cc}0 & 0 \\ 0 & 1
\end{array}\right).
\end{equation} \cite{melrosec} used the approximation $\quad r_0^2=-2 i \alpha \rho$ instead of the expansion (\ref{rad}) and
obtained the approximate high-T expressions as his final answers.

However, one can take the emergent integrals, if one considers the exact expansions
(\ref{rad},\ref{exp_t},\ref{exp_R}). Three terms appear in the expanded expression for
${\alpha^\mu}_\nu:$
\begin{mathletters}\label{calc}
\begin{equation}\label{calca}
\int^\infty_0 d\alpha \left[{t^\mu}_\nu\frac{K_2(r_0)}{r_0^2}\right],
\end{equation}
\begin{equation}\label{calcb}
\int^\infty_0 d\alpha \left[{t^\mu}_\nu\frac{K_3(r_0)\delta r^2}{r_0^3}\right],
\end{equation}
\begin{equation}\label{calcc}
\int^\infty_0 d\alpha \left[R^{\mu}\bar{R}_\nu\frac{K_3(r_0)}{r_0^3}\right].
\end{equation}
\end{mathletters} The 2-nd term (\ref{calcb}) originates from the expansion of $K_2(r)/r^2$ in $r^2$ to the first order
\begin{equation}\label{expan}
\frac{K_2(r)}{r^2}-\frac{K_2(r_0)}{r_0^2}=-\frac{\delta r^2}{2}\frac{K_3(r_0)}{r_0^3}.
\end{equation}

Integrals (\ref{calca},\ref{calcb},\ref{calcc}) can be evaluated knowing that
\begin{mathletters}\label{ident}
\begin{equation}
\int^\infty_0 d\alpha
\left[\alpha^n\frac{K_2(\sqrt{\rho^2-2i\rho\alpha})}{\rho^2-2i\rho\alpha}\right]=n!i^{n+1}\frac{K_{n-1}(\rho)}{\rho^2},
\end{equation}
\begin{equation}
\int^\infty_0 d\alpha
\left[\alpha^n\frac{K_3(\sqrt{\rho^2-2i\rho\alpha})}{(\rho^2-2i\rho\alpha)^{3/2}}\right]=n!i^{n+1}\frac{K_{n-2}(\rho)}{\rho^3}.
\end{equation}
\end{mathletters}

\subsection{Components in high-frequency limit}
I substitute the high-frequency expansions (\ref{rad},\ref{exp_t},\ref{exp_R}) into the expression
(\ref{dielectric}) for the projection of the dielectric tensor ${\varepsilon^\mu}_\nu$ with the
projection of the response tensor ${\alpha^\mu}_\nu$ (\ref{main}) and take the integrals
(\ref{calca},\ref{calcb},\ref{calcc}) analytically. The components of the dielectric tensor
(\ref{dielectric}) in the lowest orders in $\Omega_0/\omega$ are then

\begin{mathletters}\label{diag}
\begin{equation}
{\varepsilon^1}_1=1-\frac{\omega_p^2}{\omega^2}\left(\frac{K_1(\rho)}{K_2(\rho)}\left(1+\frac{\Omega_0^2}{\omega^2}
\cos^2\theta\right)+\frac{\Omega_0^2\sin^2\theta}{\omega^2\rho}\right),
\end{equation}
\begin{equation}
{\varepsilon^2}_2=1-\frac{\omega_p^2}{\omega^2}\left(\frac{K_1(\rho)}{K_2(\rho)}\left(1+\frac{\Omega_0^2}{\omega^2}\right)+\frac{7\Omega_0^2\sin^2\theta}{\omega^2\rho}
\right),
\end{equation}
\end{mathletters}
\begin{equation}\label{nondiag}
{\varepsilon^1}_2=-{\varepsilon^2}_1=-i \eta
\frac{\omega_p^2\Omega_0}{\omega^3}\frac{K_0(\rho)}{K_2(\rho)}\cos\theta,
\end{equation} where the plasma frequency $\omega_p$ in CGS units is
\begin{equation}\nonumber
\omega_p^2=\frac{4\pi n q^2}{m}.
\end{equation}

The results reproduce the non-relativistic limits for $\rho\rightarrow+\infty:$
\begin{mathletters}\label{nonrel}
\begin{equation}
{\varepsilon^1}_1=1-\frac{\omega_p^2}{\omega^2}\left(1+\frac{\Omega_0^2}{\omega^2}
\cos^2\theta\right),
\end{equation}
\begin{equation}
{\varepsilon^2}_2=1-\frac{\omega_p^2}{\omega^2}\left(1+\frac{\Omega_0^2}{\omega^2}\right),
\end{equation}
\begin{equation}
{\varepsilon^1}_2=-{\varepsilon^2}_1=-i \eta \frac{\omega_p^2\Omega_0}{\omega^3}\cos\theta,
\end{equation}
\end{mathletters} where all Bessel
functions of $\rho$ approach unity\footnote{The non-diagonal term has a wrong sign in
\cite{melrosec}.} \citep{landau10,trubnikov_book,swanson,bellan}. The corresponding relativistic
limits $\rho\rightarrow0$ of the same components are
\begin{mathletters}\label{rel}
\begin{equation}\label{rela}
{\varepsilon^1}_1=1-\frac{\omega_p^2}{\omega^2}\left(\frac1{2T}\left(1+\frac{\Omega_0^2}{\omega^2}
\cos^2\theta\right)+T\frac{\Omega_0^2\sin^2\theta}{\omega^2}\right),
\end{equation}
\begin{equation}\label{relb}
{\varepsilon^2}_2=1-\frac{\omega_p^2}{\omega^2}\left(\frac1{2T}\left(1+\frac{\Omega_0^2}{\omega^2}\right)+T\frac{7\Omega_0^2\sin^2\theta}{\omega^2}
\right),
\end{equation}
\begin{equation}\label{relc}
{\varepsilon^1}_2=-{\varepsilon^2}_1=-i \eta
\frac{\omega_p^2\Omega_0}{\omega^3}\frac{\ln(T)}{2T^2}\cos\theta,
\end{equation}
\end{mathletters}
consistent with \cite{melrosec,quataert_farad}\footnote{The diagonal plasma response is 2 times
larger in \cite{melrosec} by an error.}. The ultra-relativistic non-magnetized dispersion relation
then reads
\begin{equation}\label{disp_nonmagn} \omega^2=\frac{\omega_p^2}{2T}+c^2k^2=\frac{2\pi n q^2}{m
T}+c^2k^2
\end{equation} according to the relation (\ref{dispersion}). The expression (\ref{disp_nonmagn}) is consistent with
\cite{landau10}, chapter 32.

The plasma propagation effects can usually be described in terms of only the difference of the
diagonal components and the non-diagonal component of ${\varepsilon^\mu}_\nu.$ I define $\bf X$ to
be a vector of $T,$ $\theta,$ $\Omega_0/\omega.$ I introduce the multipliers $f({\bf X})$ and
$g({\bf X})$ to correct the expressions, when the high-frequency limit breaks. I write the
difference between the diagonal components with a multiplier $f({\bf X})$ as
\begin{equation}\label{diagdifffit}
{\varepsilon^1}_1-{\varepsilon^2}_2=f({\bf
X})\frac{\omega_p^2\Omega_0^2}{\omega^4}\left(\frac{K_{1}(T^{-1})}{K_2(T^{-1})}+6T\right)\sin^2\theta
\end{equation} and the non-diagonal component with a multiplier $g({\bf X})$ as
\begin{equation}\label{nondiagfit}
{\varepsilon^1}_2=-i \eta g({\bf
X})\frac{\omega_p^2\Omega_0}{\omega^3}\frac{K_0(T^{-1})}{K_2(T^{-1})}\cos\theta.
\end{equation} Both multipliers equal unity in the high-frequency limit $f({\bf
X})=g({\bf X})=1.$ Now we can turn to a more general case.

\subsection{Fitting formulas for higher temperatures}\label{low_dens}
\begin{figure}[h]
\plotone{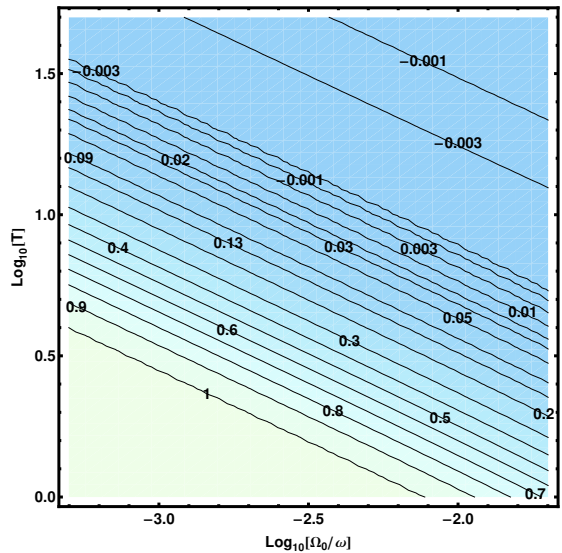} \epsscale{1}
 \caption{Multiplier $f({\bf X})$ for the difference of the diagonal components ${\varepsilon^1}_1-{\varepsilon^2}_2$ for $\theta=\pi/4.$}
 \label{fig_diag_contour}
\end{figure}

\begin{figure}[h]
\plotone{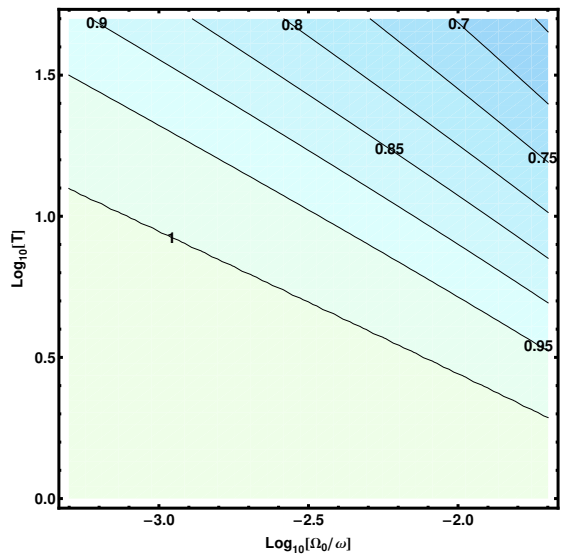} \epsscale{1}
 \caption{Multiplier $g({\bf X})$ for the non-diagonal component ${\varepsilon^1}_2$ for $\theta=\pi/4.$}
 \label{fig_nondiag_contour}
\end{figure}

The ultra-relativistic expressions (\ref{rela},\ref{relb},\ref{relc}) allow me to trace the
T-factors in front of the first 3 expansion coefficients of the dielectric tensor in $\beta.$ The
coefficient at $\beta^2$ is $\sim T^3/\ln(T)$ times larger than at $\beta.$ Thus at temperature
$T\gtrsim 10$ the 2-nd order becomes larger than the 1-st order for the ratio
$\Omega_0/\omega\sim10^{-3}.$ This indicates that the expansion in $\beta$ may become invalid at
these plasma parameters\footnote{One cannot claim that the diagonal magnetized terms become larger
then the non-diagonal \citep{melrosec}.}. The multipliers $f({\bf X})$ and $g({\bf X})$ are likely
to be far from $1.$ I consider only the real parts of these multipliers, since the imaginary parts
correspond to absorption. The contour plots of the numerically calculated $f({\bf X})$ and $g({\bf
X})$ for somewhat arbitrary $\theta=\pi/4$ are shown on Figure~\ref{fig_diag_contour} and
Figure~\ref{fig_nondiag_contour}, respectively.

Let me define $X$ to be the following combination of the parameters
\begin{equation}
X=T\sqrt{\sqrt{2}\sin\theta \left(10^3\frac{\Omega_0}{\omega}\right)}.
\end{equation}
For the fiducial $\Omega_0/\omega=10^{-3},$ $\theta=\pi/4$ the parameter $X$ is just temperature
$X=T.$

\begin{figure}[h]
\plotone{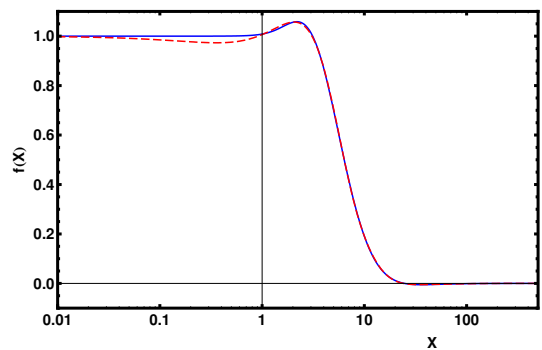} \epsscale{1}
 \caption{Multiplier $f(X)$ for the difference of the diagonal components ${\varepsilon^1}_1-{\varepsilon^2}_2.$ Dashed line --- fitting formula (\ref{gaunt_diag}).}
 \label{fig_diag_fit}
\end{figure}

\begin{figure}[h]
\plotone{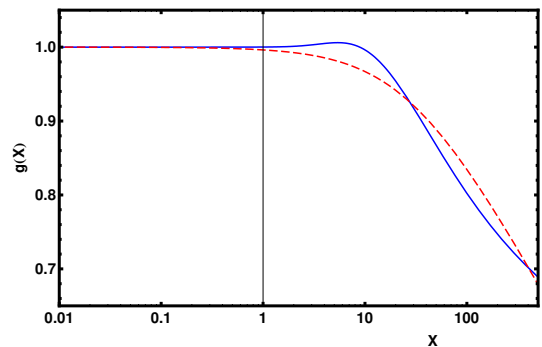} \epsscale{1}
 \caption{Multiplier $g(X)$ for the non-diagonal component ${\varepsilon^1}_2.$ Dashed line --- fitting formula (\ref{gaunt_nondiag}).}
 \label{fig_nondiag_fit}
\end{figure}

I first identify the boundaries, where the high-frequency limit is valid. Then I find a fit for the
multipliers at higher $X.$ The expression (\ref{diagdifffit}) for the difference
${\varepsilon^1}_1-{\varepsilon^2}_2$
 is accurate within $10\%$ for $X<0.1$ if we set $f(X)=1.$ The expression (\ref{nondiagfit}) for
${\varepsilon^1}_2$ is accurate within $10\%$ for $X<30$ if we set $g(X)=1.$ The accuracy depends
on the parameter $X$ rather than on the individual parameters $T,$ $\Omega_0/\omega,$ $\theta.$ The
expression
\begin{eqnarray}\label{gaunt_diag}
f(X)=2.011\exp\left(-\frac{X^{1.035}}{4.7}\right)-\nonumber\\
-\cos\left(\frac{X}2\right)\exp\left(-\frac{X^{1.2}}{2.73}\right)-0.011\exp\left(-\frac{X}{47.2}\right)
\end{eqnarray}
extends the applicability domain of the formula (\ref{diagdifffit}) up to $X\sim 200.$
Figure~\ref{fig_diag_fit} shows the fit for $f(X)$ in comparison with the numerical results.
The expression
\begin{equation}\label{gaunt_nondiag}
g(X)=1-0.11\ln(1+0.035X)
\end{equation} extends up to $X \sim 200$ the domain of the formula (\ref{nondiagfit}). Figure~\ref{fig_nondiag_fit} shows the fit for $g(X)$
in comparison with the numerical results.

\subsection{Exact plasma response}\label{exact}
The expression for the response tensor (\ref{main}) is written for a vacuum wave with $|{\bf
k}|c=\omega.$ In the real plasma, the wave is modified by the plasma response. A more general
self-consistent response tensor should be used \citep{trubnikov,melrosec}. One needs to solve a
dispersion relation similar to the relation (\ref{dispersion}) to obtain the eigenmodes. Thus the
eigenmodes and the response tensor should be computed self-consistently. One should not forget
about the antihermitian and longitudinal components of the dielectric tensor
${\varepsilon^\mu}_\nu$ that modify the dispersion relation.

\subsection{Eigenmodes}\label{modes} The above calculation is applicable also to a non-magnetized plasma.
Dispersion relation of EM waves in a non-magnetized plasma reads
\begin{equation}\label{nonmagn}
\omega^2=k^2 c^2+\omega^2_p \frac{K_1(T^{-1})}{K_2(T^{-1})}
\end{equation} in a high-frequency approximation $\omega\gg \omega_p$. The opposite limit of $k c \ll \omega$
was considered by \cite{bergman}.

\begin{figure}[h]
\vspace*{-20mm} \plotone{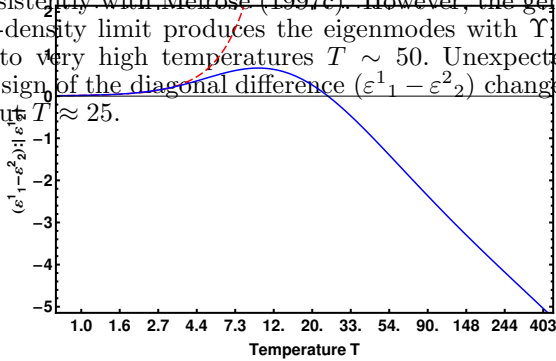} \epsscale{1}
 \caption{Ellipticity $\Upsilon=({\varepsilon^1}_1-{\varepsilon^2}_2):|{\varepsilon^1}_2|$ of eigenmodes.
  The absolute value of the ratio $\Upsilon$ much above unity --- linear eigenmodes, much below unity --- circular eigenmodes.
  Solid line --- this paper, dashed line --- previous calculations.}
 \label{fig_ratio_fit}
\end{figure}

Now we turn to the magnetized case. \cite{melrosec} only considered the first terms of
 in the expansion of ${\alpha^\mu}_\nu$ in $\beta$ to get the eigenmodes. I do the next step: consider the full expression in $\beta$ in the
low-density regime $k c=\omega,$ but consider only the hermitian part of ${\alpha^\mu}_\nu$ in
computations. The ellipticity $\Upsilon=({\varepsilon^1}_1-{\varepsilon^2}_2):|{\varepsilon^1}_2|$
determines the type of eigenmodes. If $|\Upsilon|\gg 1,$ then the eigenmodes are linearly polarized
unless $\theta$ is close to $0.$ If $|\Upsilon| \ll 1,$ then the eigenmodes are circularly
polarized for $\theta$ far from $\pi/2.$ Let me consider the fiducial model with
$\Omega_0/\omega=10^{-3}$ and $\theta=\pi/4.$ Figure~\ref{fig_ratio_fit} shows the ratio $\Upsilon$
calculated in a high-frequency approximation (see \S~\ref{high_freq}) (dashed line) and in a
general low-density approximation (see \S~\ref{low_dens}) (solid line). The high-frequency
approximation produces the linear eigenmodes already at $T\gtrsim10$ consistently with
\cite{melrosec}. However, the general low-density limit produces the eigenmodes with
$\Upsilon\sim1$ up to very high temperatures $T\sim 50.$ Unexpectedly, the sign of the diagonal
difference $({\varepsilon^1}_1-{\varepsilon^2}_2)$ changes at about $T\approx25.$

\section{Applications}\label{applications}
The calculated transrelativistic propagation effects have far-reaching consequences in many topics
of astronomy. Let me concentrate on four applications: propagation delay, Faraday rotation measure
of light from the Galactic Center (GC), circularly polarized light from the GC, diagnostics of
jets.
\subsection{Dispersion measure}
Propagation delay is an important effect in pulsar dispersion \citep{phillips}. The relativistic
part of this delay can be obtained from the dispersion relation (\ref{nonmagn}). I retain only the
first-order correction in $T,$ since $T\ll 1$ in the interstellar medium \citep{cox}. Since
$K_1(T^{-1})/K_2(T^{-1})\approx 1-3T/2$ at low $T,$ the non-relativistic Dispersion Measure (DM)
should be modified as
\begin{equation}\label{DM}
{DM_{\rm rel}=DM_{\rm nonrel}}\left(1-\frac32T\right).
\end{equation} This shows that the gas density is slightly underestimated, if the non-relativistic formulas are used\footnote{The formula in \cite{phillips}
has no references/checks and is not correct.}. However, the relativistic correction to the DM is
small and can be neglected in most practical cases when $T \ll 1.$ The effects in magnetized plasma
are also relevant for pulsars.

\subsection{Magnetized radiative transfer}
\subsubsection{General formulae} Relativistic plasmas exhibit a generalized Faraday rotation for a general orientation of the
magnetic field \citep{mueller}. One can decompose it into two effects: Faraday rotation and Faraday
conversion. The former operates alone at $\theta=0,\pi,$ the latter operates alone at
$\theta=\pi/2,$ and both should be considered together for the intermediate angles. The transfer
equations (Mueller calculus) for the Stokes parameters $I,$ $Q,$ $U,$ $V$ were devised to treat
together the propagation effects, emission, and absorption \citep{mueller,melrose_dispersive}. Good
approximations for emission and absorption have been long known
\citep{trubnikov,rybicki,melrose_dispersive,wolfe}. Now one can combine them with the proper
approximations of the propagation effects given by
\begin{equation}\label{transfer}
\frac{d}{ds}\left(\begin{array}{c}
  I \\  Q \\  U \\  V \\
\end{array}\right)=\left(%
\begin{array}{cccc}
  0 & 0 & 0 & 0 \\
  0 & 0 & -\rho_V & \rho_U \\
  0 & \rho_V & 0 & -\rho_Q \\
  0 & -\rho_U & \rho_Q & 0 \\
\end{array}%
\right)\left(\begin{array}{c}
  I \\  Q \\  U \\  V \\
\end{array}\right),
\end{equation}
\begin{equation}
\rho_V=-\frac{\omega}{c} i {\varepsilon^1}_2, \quad
\rho_Q=-\frac{\omega}{2c}({\varepsilon^1}_1-{\varepsilon^2}_2), \quad \rho_U=0,
\end{equation} and do the radiative transfer calculations. Here ${\varepsilon^\mu}_\nu$ stands for the Hermitean
part given by the relations (\ref{gaunt_diag},\ref{gaunt_nondiag}) with the real multipliers $f(X)$
and $g(X).$ One of the most interesting objects for such calculations is our Galactic Center Sgr
A*.

The transfer equations were recently solved for a simple time-independent dynamical model of the GC
accretion \citep{huang}. The authors treat the ordinary and extraordinary modes as linearly
polarized. They assume these eigenmodes constitute a basis, where either $U$ or $Q$ components of
emissivity and propagation coefficients vanish. Actually, $U$ components vanish ($\rho_U=0$)
already in the basis $({\bf e}^1, {\bf e}^2),$ since the projection of the magnetic field onto
$({\bf e}^1, {\bf e}^2)$ is parallel to ${\bf e}^1$ (see \cite{melrose_dispersive} p.184). As I
have shown in the \S~\ref{modes}, plasma modes are far from being linearly polarized at
temperatures $T\lesssim10$ estimated for the GC \citep{sharma_T}. Thus, the propagation
coefficients should be taken from equations (\ref{diagdifffit}) and (\ref{nondiagfit}). The Faraday
conversion coefficient $\rho_Q$ cannot be defined via emissivities and Faraday rotation coefficient
$\rho_V$ as in \cite{huang}. The Faraday rotation measure was calculated from a simulated accretion
profile in \cite{sharma_farad}. However, the paper considered only the Faraday rotation and did not
carry out the self-consistent treatment of propagation. It is impossible to disentangle the effects
of Faraday rotation and Faraday conversion in a relativistic plasma.

\subsubsection{Faraday rotation} 
\begin{figure}[h]
\plotone{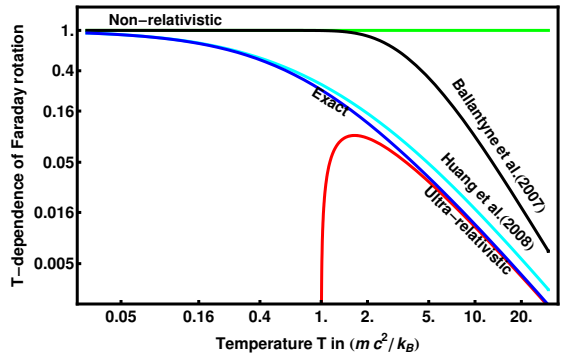} \epsscale{1}
 \caption{Temperature dependence of the Faraday rotation measure.}
 \label{fig_mildly_bridge}
\end{figure}

The crucial part of any radiative transfer is the proper transfer coefficients. It allows one to
estimate the electron density near the accreting object \citep{quataert_farad,shcher}. Several
formulas were suggested for the temperature dependence of the component ${\varepsilon^1}_2$
responsible for Faraday rotation. These formulas were yet given for the high-frequency
approximation (see \S~\ref{high_freq}). Let me compare them with the exact temperature dependence
(\ref{nondiag}) $J=K_0(T^{-1})/K_2(T^{-1})$ and its limits. The limits are $J\rightarrow1$ as
$T\rightarrow0$ and $J\rightarrow \ln(T)/(2T^2)$ as $T\rightarrow+\infty.$ The results
 of this comparison are shown on Figure~\ref{fig_mildly_bridge}.

 \cite{ballantyne}\footnote{The paper \cite{ballantyne} has likely confused the 3-D projection of the 4-D
response tensor in $j^\mu=\alpha^{\mu\nu}A_\nu$ \citep{melrosec} with the 3-D response tensor ${\bf
j}=\alpha_{i j}{\bf A}$ that has the opposite sign.}  divided the thermal distribution
 into ultra-relativistic and non-relativistic parts as marked by the electron energy $\gamma_{\rm crit}=10.$
They sum the contributions of both species with calculated densities. To make a plot, I take their
effective temperature $\Theta$ of plasma above $\gamma_{\rm crit}$ to be just temperature
$\Theta=T$ and not the average kinetic energy as \cite{ballantyne} suggest. This brings $\Theta$ to
lower values and decreases the rotation measure. Even with this decrease the rotation measure is
severely overestimated at $T\sim1.$ The convergence to the relativistic limit is not achieved even
at $T\sim30.$ The paper \cite{huang} found the simpler fitting formula that reproduces the limits.
Their expression is quite accurate.\footnote{"Temperature" $\gamma_c$ in \cite{huang} should be
redefined as $\gamma_c=1+T$, otherwise the lower limit is not reproduced.}

\subsubsection{Faraday conversion}The increase in the circular polarization of Sgr A* at frequency 1THz is predicted by
\cite{huang}. The phase of Faraday conversion approaches unity and the destructive interference
does not occur at this frequency. The result seems to be qualitatively correct regardless of the
expression for the conversion measure, but the proper expressions (\ref{diagdifffit}) and
(\ref{nondiagfit}) should be used for quantitative predictions.

\subsubsection{Jets}The better treatment of propagation effects may also play a role in observations of jets. As
we saw in \S~\ref{low_dens}, the propagation effects in thermal plasma cannot be described in the
lowest orders in $\Omega_0/\omega,$ if the temperature $T$ is sufficiently high. Power-law
distribution of electrons can have a quite high effective temperature. Thus the high-frequency
limit \citep{sazonov,jones,melroseb} may not approximate well the hermitian part of the response
tensor. Careful analysis of jet observations \citep{beckert,wardle} may be needed. It should be
based at least on the expressions for ${\varepsilon^\mu}_\nu$ in a general low-density regime.

\section{Discussion \& Conclusion}\label{discussion}
This paper presents several new calculations and amends the previous calculations of propagation
effects in uniform magnetized plasma with thermal particle distribution equation (\ref{distrib}).
The expression (\ref{main}) for the correct response tensor is given in a high-frequency
approximation. The exact temperature dependence (\ref{diag}) and (\ref{nondiag}) is found in first
orders in $\Omega_0/\omega$ in addition to the known highly-relativistic and non-relativistic
results. The higher order terms may be important for relativistic plasmas in jets and hot accretion
flows. The fitting expressions (\ref{gaunt_diag}) and (\ref{gaunt_nondiag}) are found for the
dielectric tensor components (\ref{diagdifffit}) and (\ref{nondiagfit}) at relatively high
temperatures.

The results of numerical computations are given only when the corresponding analytical formulas are
found. One can always compute the needed coefficients numerically for every particular frequency
$\omega$, plasma frequency $\omega_p,$ cyclotron frequency $\Omega_0,$ and distribution of
electrons. However, the analytic formulas offer a simpler and faster way of dealing with the
radiative transfer for a non-specialist. The eigenmodes were not considered in much detail, since
radiative transfer problems do not require a knowledge of eigenmodes. However the knowledge of
eigenmodes is needed to compute the self-consistent response tensor (see \S~\ref{exact}).

The response tensor in the form (\ref{main}) can be expanded in $\Omega_0/\omega$ and
$\omega_p/\omega.$ This expansion is of mathematical interest and will be presented in a subsequent
paper as well as the expressions for a power-law electron distribution. Propagation through
non-magnetized plasmas will also be considered separately.

\acknowledgements The author is grateful to Ramesh Narayan for fruitful discussions and Diego Munoz
for pointing out relevant references. I thank the anonymous referee for helpful suggestions that
improved the paper.

\end{document}